\documentclass{article}
\usepackage{spconfa4,amsmath,graphicx}
\usepackage{acronym}
\usepackage{tikz}
\usepackage{amssymb}
\usepackage{caption}
\usepackage{xcolor}
\definecolor{matplotlibgreen}{rgb}{0, 0.5, 0}
\definecolor{matplotliborange}{rgb}{1.0, 0.5, 0.0}
\definecolor{matplotlibblue}{rgb}{0.0, 0.447, 0.741}
\definecolor{matplotlibred}{rgb}{0.839, 0.153, 0.157}
\title{Directivity-Conditioned Low-Latency Neural Filtering for Speech Enhancement in Hearing Aids}
\name{Lennart Uphaus$^1$, André Merboldt$^2$ , Markus Hofbauer$^3$ , Timo Gerkmann$^1$\thanks{For this work the HPC-cluster Hummel-2 at the University of Hamburg was used. The cluster was funded by Deutsche Forschungsgemeinschaft (DFG, German Research Foundation) – 498394658.}}
\address{
    $^1$ Signal Processing (SP), University of Hamburg, Germany \\
    $^2$ Audatic GmbH, Berlin, Germany 
    $^3$ Sonova, Stäfa, Switzerland 
}

\usepackage[acronym]{glossaries}
\makeglossaries
\newacronym{DNN}{DNN}{deep neural network}
\newacronym{FiLM}{FiLM}{feature-wise linear modulation}
\newacronym{MVDR}{MVDR}{minimum-variance distortionless response}
\newacronym{DoA}{DoA}{direction-of-arrival}
\newacronym{STFT}{STFT}{short-time Fourier transform}
\newacronym{SNR}{SNR}{signal-to-noise ratio}
\newacronym{NDF}{NDF}{neural directional filtering}
\newacronym{RNN}{RNN}{recurrent neural network}
\newacronym{HRTF}{HRTF}{head-related transfer function}
\newacronym{BTE}{BTE}{behind-the-ear}
\newacronym{RIR}{RIR}{room impulse response}
\newacronym{FiLM-OSN}{FiLM-OSN}{FiLM-OnlineSpatialNet}
\newacronym{FT-JNF}{FT-JNF}{frequency-time joint nonlinear filter}
\newacronym{OSN}{OSN}{OnlineSpatialNet}
\newacronym{iSTFT}{iSTFT}{inverse short-time Fourier transform}
\newacronym{FiLM-JNF}{FiLM-JNF}{FiLM Joint Nonlinear Filter}
\newacronym{BRIR}{BRIR}{binaural room impulse response}
\newacronym{HARTF}{HARTF}{Hearing Aid Related Transfer Function}
\newacronym{WSJ0}{WSJ0}{Wall Street Journal}
\newacronym{IPD}{IPD}{interaural phase difference}
\newacronym{PESQ}{PESQ}{Perceptual Evaluation of Speech Quality}
\newacronym{SI-SDR}{SI-SDR}{scale-invariant signal-to-distortion ratio}
\newacronym{ESTOI}{ESTOI}{extended short-time objective intelligibility}
\newacronym{LSTM}{LSTM}{long short-term memory}
\newacronym{ITD}{ITD}{interaural time difference}
\begin{document}

\maketitle
\begin{abstract}
Latest advances in neural directional filtering show exceptional results in adapting the direction and shape of directivity patterns during the inference phase. However, in the existing methods for adapting directivity patterns during inference, important real-world constraints have been disregarded. Particularly for hearing devices, scenarios are often much more dynamic, microphone positions vary with head diameter and hearing aid placement, head-shadow effects occur, and strict latency constraints apply.

In this work, we propose a novel low-latency (10~ms) \gls{DNN} taking the above requirements of hearing devices into account. As in recent work, we use \gls{FiLM} to steer the directivity patterns during testing. To preserve the desired directivity pattern, a loss function is proposed that maintains the spectral cross-channel relationships. Interestingly, we are able to achieve similar results to methods with relaxed latency constraints. 
\end{abstract}
\section{Introduction}
\label{sec:intro}
Due to the cocktail-party effect, normal hearing people are able to follow conversations with multiple people and background noise easily by dynamically shifting the attention to the desired source \cite{Cherry1953}. However, this changes for hearing-impaired people. Common speech enhancement algorithms concentrate on the preservation of a single target speaker while interfering sources are attenuated \cite{Pandey2022, Westhausen2024}. But especially in hearing device scenarios with multiple or changing targets, other objectives are required. A complete suppression of interfering noises is not preferred, but merely a reduction, such that an awareness of the spatial situation is maintained. Furthermore, an adaptable directivity pattern is desirable for more effective signal enhancement in dynamic multi-source scenarios.
Classical approaches such as \gls{MVDR} beamformer depend on the estimation of the steering vector \cite{Habets2010}. Since the \gls{DoA} estimation can be challenging in low \glspl{SNR} with non-speech signals, the application of such a method is therefore disadvantageous \cite{Thiergart2012}. 

As a possible solution, many \gls{NDF} approaches have been released in recent years \cite{Tesch2023, Wechsler2024, Gu2024, Huang2025}. While in \gls{FT-JNF} \cite{Tesch2024} the \gls{NDF} is conditioned on a target direction, in \cite{Gu2024} specified zones can be extracted by using region queries (e.g angular width, distance or shape).
An extension of the \gls{FT-JNF} \cite{Tesch2024}, called the \gls{FiLM-JNF}, is proposed in \cite{Huang2025}. It uses a conditioning layer to train a \gls{DNN} directly on different directivity patterns that can be adjusted during inference. Both approaches employ circular arrays and produce an enhanced single-channel speech estimate. 

Another challenge when developing hearing device algorithms is the constraint regarding latency ($\le 10~$ms) \cite{Stone2003}. The total latency in this paper consists of algorithmic latency and processing latency. For an STFT-based algorithm the algorithmic latency is given by the STFT synthesis window length. The processing latency is the time required to process one signal segment. For real-time processing, this must be lower or equal to the STFT hop size \cite{welker2025}.
To reduce the algorithmic latency, shorter \gls{STFT} windows can be employed, which results in lower frequency resolution and may deteriorate the performance. Many low-latency approaches for speech enhancement and separation have been proposed and investigated recently \cite{PandeyRNN2023, Mesgarani2019, Wu2025}. Nevertheless the recent \gls{NDF} approaches exhibit higher total latencies of 40~ms to 50~ms that are prohibitive for hearing devices. 

This article examines how reducing algorithmic latency affects the performance of \gls{FiLM-JNF} in terms of speech quality and directivity pattern estimation. We propose a low-latency \gls{NDF} approach that leverages a \gls{FiLM} layer based on the findings in \cite{Huang2025}, which also serves as our baseline. To tackle more realistic environments, moderate RT$_\text{60}$ of $0.2-0.5$~s are considered. Furthermore, we show that phase regularization in the loss function is essential for accurately reconstructing the model’s directivity pattern.

\section{Problem definition}
\label{Problem definition}
We consider an acoustic scene with $P$ speakers surrounding a head with \gls{BTE} hearing devices in a reverberant environment. Since most available datasets for \gls{BTE} hearing devices \cite{Pausch2022} consist of two microphones per device, we use a microphone array with a total of $Q = 4$ channels. The impulse response between the $p$-th speaker's position and the $q$-th microphone can be disentangled into the direct path $h_{q,p}^{\mathrm{direct}}$ and the reverberant path $h_{q,p}^{\mathrm{reverb}}$. In the time domain, the $q$-th microphone signal can be written as  
\begin{align}    
    y_q(t) &= \sum_{p = 1}^P s_{p}(t) \ast (h_{q, p}^{\mathrm{direct}}(t) + h_{q,p}^{\mathrm{reverb}}(t))\\
     &= \sum_{p = 1}^P (x_{q,p}(t) + v_{q,p}(t)),
\end{align}
with the convolution operator $\ast$, the direct-path signal $x_{q,p} = s_p \ast h_{q,p}^{\mathrm{direct}}$ and reverberation $v_{q,p} = s_p\ast h_{q,p}^{\mathrm{reverb}}$.
We aim to reconstruct an anechoic binaural speech signal that follows the attenuation constraints given by a defined directivity pattern $\Lambda_t(\theta)$ at time index $t$, which depends on the \gls{DoA} angle $\theta$ of the $p$-th speaker relative to the microphone array. This is given by 
\begin{equation}
    z_q(t) = \sum_{p = 1}^P \Lambda_t(\theta_p)x_{q,p}(t) \quad q = \{0,2\}.
\end{equation}
\section{Proposed method}
\subsection{Low-latency architecture}
The approach of \cite{Huang2025} shows exceptional results in the field of neural directional beamforming based on the \gls{FT-JNF} architecture proposed in \cite{Tesch2024}. We found that reducing the algorithmic latency by taking short 8~ms STFT segments in the \gls{FT-JNF} architecture significantly decreases the performance. Therefore, we propose the \gls{FiLM-OSN}, in which we extend the \gls{OSN} \cite{QuanOnSpatial2024} with the \gls{FiLM} mechanism \cite{Perez2018} to ensure directional conditioning while increasing speech quality in a low-latency setup (see Figure \ref{fig:network_architecture}). 

\usetikzlibrary{positioning}
\usetikzlibrary{decorations.pathreplacing}
\tikzset{
    architecture_pic/.pic = {

    \node (input) at (0, 1.25) {$y(t)$};
    \node (STFT) at (0,0) [draw,thick,minimum width=1.5cm,minimum height=0.25cm] {STFT};

    \node (Tconv) at (0, -1.25) [draw, thick, minimum width = 3.5cm, minimum height = 0.25cm, fill = gray!40] {T-Conv-1d};
    \node (CrossBand) at (0, -2.5) [draw, thick, minimum width = 3.5cm, minimum height=0.25cm, fill = gray!40] {Cross-band Block};

    \node (FiLM) at (0, -3.75) [draw, thick, minimum width = 3.5cm, minimum height=0.25cm, fill = green!80] {FiLM};

    \node (NarrowBand) at (0, -5.0) [draw, thick, minimum width = 3.5cm, minimum height = 0.25cm, fill = gray!40] {Narrow-band Block};

    \node (Linear) at (0, -6.25) [draw, thick, minimum width = 3.5cm, minimum height = 0.25cm, fill = gray!40] {Linear};

    \node (iSTFT) at (0, -7.5) [draw, thick, minimum width = 1.5cm, minimum height = 0.25cm] {iSTFT};

     \node (output) at (0, -8.75) {$\hat{s}_q(t)$};

    \draw[->] (input) --(STFT);
    \node[above=0.1pt of STFT, xshift=10mm] (input_size)
    {[B, F, T, 2Q]};

    \draw[->] (STFT) --(Tconv);
    \node[above=0.1pt of Tconv, xshift=10mm] (input_tconv)
    {[B, F, T$\times$2Q]};
    
    \draw[->] (Tconv) -- (CrossBand);
    \node[above=0.1pt of CrossBand, xshift=10mm] (input_cross)
    {[B, T, F$\times$C]};
    
    \draw[->] (CrossBand) -- (FiLM);
    \node[above = 0.1pt of FiLM, xshift=10mm] (input_film)
    {[B, F, T, C]};
    
    \draw[->] (FiLM) -- (NarrowBand);
    \node[above=0.1pt of NarrowBand, xshift=10mm] (input_narrow)
    {[B,F, T$\times $C]};
    \draw[->] (NarrowBand) -- (Linear);
    \node[above=0.1pt of Linear, xshift=10mm] (input_linear)
    {[B,F, T, C]};
    
    \draw[->] (Linear) -- (iSTFT);
    \node[above=0.1pt of iSTFT, xshift=10mm] (input_istft)
    {[B,F, T, 4]};
    \node[right = of FiLM] (Pattern) [draw, thick, minimum width = 2.5cm, minimum height = 0.15cm, fill = green!80] {Pattern input};

    \draw[->] (Pattern) -- (FiLM);

    \draw[->] (iSTFT) -- (output);
    \draw [decorate,decoration={brace,amplitude=5pt,mirror,raise=4ex}]
  (-1.5,-2.4) -- (-1.5,-5.2) node[midway, xshift = -1.0cm,rotate = 90]{L times};

    \node (img) at (5.0,0) {\includegraphics[width = 3cm]{beampattern.png}};

    \draw[->, shorten <= -0.31cm] (img) -- (Pattern);
    
}
}

\begin{figure}[ht]
    \centering
    \begin{tikzpicture}
        
    \node at (0,0) {\includegraphics[width = 0.75\linewidth]{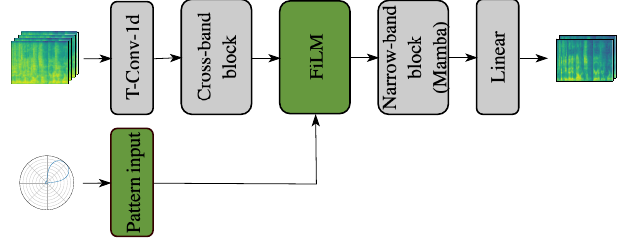}};
    \draw [decorate,decoration={brace,amplitude=5pt,raise=4ex}]
  (-1.17,0.7) -- (1.20,0.7) node[midway,yshift=2.7em]{\small{$L$ times}};
    \end{tikzpicture}
    \caption{\gls{FiLM-OSN} architecture for neural directional beamforming.}
    
    \label{fig:network_architecture}
\end{figure}
Complex \gls{STFT} coefficients of the input mixture are concatenated, as
\begin{equation}
    Y_{\text{in}} = [\Re\{Y_1, Y_2,...,Y_Q\}, \Im\{Y_1, Y_2,...,Y_Q\}]\in \mathbb{R}^{2Q}
\end{equation}
and serve as inputs to the \gls{FiLM-OSN} network. Here, $\Re\{\cdot\}$ and $\Im\{\cdot\}$ denote the real and imaginary part.
The core of the \gls{FiLM-OSN} architecture consists of $L$ interleaved cross-band blocks, \gls{FiLM}, and narrow-band blocks, each with an output dimension of $F\times T \times C$, where $C = 96$. The cross-band blocks facilitate spectral and spatial information processing through two frequency-convolution modules and one full-band linear module, operating independently across time frames. The \gls{FiLM} layer employs linear mapping from the directivity pattern vector to the channel size $C$, thereby conditioning the narrow-band block toward the desired pattern. The narrow-band blocks integrate a state-space model (Mamba) \cite{Gu2023} and a time-convolutional module, which process each frequency bin independently. The channel dimension of the time-convolutional module is increased to $C' = 196$. The two channels output of the \gls{FiLM-OSN} is produced via a linear layer and represents the predicted \gls{STFT} coefficients, which are transformed by the \gls{iSTFT} to reconstruct the estimated time-domain target signal $\hat{z}_{left}$ and $\hat{z}_{right}$. Further details about the causal implementation as well as the base network architecture can be found in \cite{QuanOnSpatial2024} and \cite{QuanSpatial2024}.

When considering low-latency approaches, convolutional layers play a significant role \cite{Mesgarani2019}. U-Net structures are also utilized, where encoders and decoders are characterized by convolutional layers \cite{Wang2023}. As for \gls{FT-JNF}, the lower frequency resolution consequently results in a shorter sequence length for the spectral \gls{LSTM}, which could affect performance. In contrast to the temporal \gls{LSTM} layer for narrow-band processing in \gls{FT-JNF}, a dynamic state-space model (Mamba) is used in \gls{OSN}, which can model long-term dependencies better. 

During training, we employ the batch-aggregated normalized $\mathcal{L}_1$-loss shown in \cite{Huang2025}, which is defined as
\begin{equation}
    \mathcal{L}_1
=
\frac{\sum_{b=1}^{B} \left\lVert z_{l,r}^{b} - \hat{z}_{l,r}^{\,b} \right\rVert_{1}}
{\sum_{b=1}^{B} \left\lVert z_{l,r}^{b} \right\rVert_{1} + \epsilon},
\end{equation}
where $\epsilon$ is a small constant value, B represents the batch-size and the indices $l,r$ indicate the binaural signal.
To promote convergence to the desired directivity pattern, we adapt the loss function by adding a phase preservation term that penalizes \gls{IPD} deviations between the target and the prediction, as given by
\begin{equation}
\mathcal{L}_{\mathrm{IPD}} =
\frac{
\sum_{f,t}
\left| Y_{0}(f,t) \right|^{2}
\left[
1 - \cos\!\left(
\hat{\Phi}(f,t) - \Phi(f,t)
\right)
\right]
}{
\sum_{f,t}
\left| Y_{0}(f,t) \right|^{2} + \epsilon
},
\end{equation}
where $\hat{\Phi}(f,t)$ and $\Phi(f,t)$ are the cross-channel phase differences of the prediction and the target. We make use of the cosine to avoid phase ambiguities \cite{Wiley1999}. To reduce the contribution of the low-energy components, we use $Y_0$ as a weighting factor. A loss weighting $\alpha = 0.03$ is applied, which leads to
\begin{equation}
    \mathcal{L}_{1,\text{IPD}} = \mathcal{L}_1 + \alpha \mathcal{L}_{\text{IPD}}.
\end{equation}
\subsection{Directivity pattern strategy}
Recent methods \cite{Wechsler2024, Huang2025} use differential microphone array (DMA) patterns as training objectives. However, the width of the main-lobe for these patterns cannot be precisely specified by an angular width. Therefore, we propose a cosine-based directivity pattern with a configurable main-lobe width $W$. To ensure spatial awareness for the user, we limit the maximum attenuation. The directivity pattern is defined as
\begin{equation}
    \Lambda(\theta) = \begin{cases}
        |\cos(\frac{\pi}{W} \angle e^{j(\theta - \theta_d)})|, \quad{\text{if } |\angle e^{j(\theta - \theta_d)}| \le \frac{W}{2}}\\
        M, \qquad \qquad \quad \quad \text{  else}
        \end{cases} 
\end{equation}

where $M$ is the attenuation limit. $\theta_d$ represents the orientation of the directivity pattern relative to the head orientation.
\tikzset{
    head_pic/.pic = {
    \draw[black] (2.0, 2.0) circle (0.25);

            \filldraw[blue] (2.05, 2.25) circle (0.03); 

            \filldraw[blue] (1.95, 2.25) circle (0.03); 
            \filldraw[blue] (1.95, 1.75) circle (0.03);%
            \filldraw[blue] (2.05, 1.75) circle (0.03); %

            \draw[] (2.25,1.95) -- (2.35, 2.00);
            \draw[] (2.35, 2.00) -- (2.25, 2.05);
    }
    }
\tikzset{
    geometry_pic/.pic = {

\def\W{8}            %
\def\H{5}            %
\def\Rhead{0.3}      %
\def\RayLen{20}      %

\path[fill=gray!15,draw=black] 
    (-\W/2,-\H/2) rectangle (\W/2,\H/2);

\begin{scope}
  \clip (-\W/2,-\H/2) rectangle (\W/2,\H/2);

  \foreach \a in {90,18,-54,-126,162}{
    \draw[dashed] (0,0) -- ++(\a:\RayLen);
  }

  \foreach \p in {( -1.5,  1.3),
                 (  1.5,  1.8),
                 (  1.3,  -0.3),
                 ( -1.4, -0.8),
                 (  0.5, -1.8)}{
    \fill[orange] \p circle[radius=0.10];
  }
\end{scope}

\pic at (2.4,-2.4) [scale = 1.2, rotate around={90:(0,0)}, transform shape](0,0){head_pic};}}
\section{Experiments}
\subsection{Dataset}
For training and testing purposes, we simulate rooms consisting of $P = 5$ speakers. As a speech dataset, we use the \gls{WSJ0} corpus. The impulse responses for simulating the binaural signal captured by a \gls{BTE} hearing device are provided by the dataset in \cite{Denk2018}. It comprises a 3-channel microphone configuration per hearing device. To meet our requirements for a dual-microphone configuration, we omit the middle microphone. The dataset contains 19 \gls{HARTF} files, including four KEMAR heads and 15 \gls{HARTF}s from individual subjects. 13, 5 and 2 \gls{HARTF}s are used for training, validation and testing, respectively. All sets contain both individual heads as well as at least one KEMAR head.
The speakers are evenly distributed around the microphone array. For this, the room is divided into areas of 72° angular width. Per area, one speaker is placed randomly with a distance of 1.20$\pm$0.20~m relative to the microphone array, where the minimum angular distance between speakers is set to 10°. Room characteristics, as well as a visualization of a room sample, are given in Figure \ref{fig:dataset_setup}.
\begin{figure}[ht]
    \begin{minipage}{0.55\linewidth}

    \begin{tikzpicture}
        \pic at (0,0) [scale = 0.5]{geometry_pic};
    \end{tikzpicture}

    \end{minipage}
    \begin{minipage}{0.35\linewidth}
        \begin{tabular}{ll}
        \hline
        \multicolumn{2}{c}{\textbf{Room characteristics}} \\\hline\hline
        Width  &$3 - 9$~m\\
        Length   & $2.5 - 5$~m\\
        Height & $2.2 - 3.5$~m\\
        $T_{60}$ & $0.2 - 0.5$~s\\ 
        \hline
        \end{tabular}
    \label{tab:room_dims}
    \end{minipage}
    \captionof{figure}{Visualization of the dataset setup together with important room characteristics. The orange and blue dots represent the sources and the BTE hearing device setup.}
    \label{fig:dataset_setup}
\end{figure}
While the mouths of the speakers are set to a height of 1.60~m with a standard deviation of 0.08, the microphone array is fixed at 1.5~m. The selection of the location and orientation of the listener's head is random for each room, ensuring that it is at least 1.2~m away from the walls.
Each source is set to a random loudness between -25~LUFS and -20~LUFS. All in all, 6000, 1200 and 600 \glspl{BRIR} are simulated for training, validation and testing. Audio examples can be found on our webpage\footnote{https://sp-uhh.github.io/film-osn/}.
\subsection{Network architectures and training details}
As our baseline, we use the \gls{FiLM-JNF}.
We adapt the spectral mask to output two channels instead of only one to produce a binaural signal. Throughout the experiments, $\sqrt{\text{Hann}}$ windows with two different \gls{STFT} window configurations are used. As an upper limit, we utilize a length of 32~ms with a hop size of 8~ms. To meet the low-latency constraints, the window size is reduced to 8~ms with a hop size of 2~ms. The employed window sizes are indicated in the network index for clarity. 

\textbf{\textit{FiLM-OnlineSpatialNet}}
Our proposed \gls{FiLM-OSN}$_\text{8ms}$ is trained with a \gls{STFT} window size of 8~ms and a hop size of 2~ms, which results in a total latency of 10~ms. We utilize the parameters of the SpatialNet-small in \cite{QuanSpatial2024} while using only $L = 4$ for complexity reduction and faster training process. The \gls{FiLM} layer output size is reduced from 512 to 96 channels to align it with \gls{OSN} configurations.
\par 
\textbf{\textit{Training configuration}}
The used directivity patterns are restricted to the positions of the sources relative to the microphone array of each sample. This ensures that at least one source in the target signal is not attenuated. This restriction leads to five possible directivity patterns per dataset sample, which results in $6000 \times5$, $1200\times5$ and $600\times5$ samples for training, validation and testing. 
The directivity pattern is represented by a 72-dimensional vector sampled at 5° intervals and used as input to the FiLM layer. For training and testing, we employ 3-second speech signals.
The maximum number of epochs is set to 100, with the model being early-stopped if the validation loss does not decrease for 8 consecutive epochs. For the \gls{FiLM-JNF}, we use a scheduler that reduces the learning rate by a factor of 0.5 when the validation loss plateaus for 3 epochs. For the \gls{FiLM-OSN}, a scheduler with exponential learning rate decay is used, with a decay of $\gamma = 0.99$. Batch size and learning rate are set to 6 and $10^{-3}$. The low-latency \gls{FiLM-JNF} is trained with smaller learning rate ($10^{-4}$) to avoid instabilities. $M$, and $W$ are set to $-20$~dB and 90°.
\section{Results}
As evaluation metrics, we employ \gls{SI-SDR}, \gls{ESTOI} and \gls{PESQ} \cite{RixPESQ2001} to cover distortions, speech intelligibility and perceptual quality. To compute the directivity patterns, 50 speech signals from the test set are spatialized at three-degree intervals via the KEMAR \gls{HARTF} of the test set with no reverberation. The pattern is then computed via the wide-band power ratio $\xi$, defined as
\begin{equation}
    \xi(\theta)
=\frac{
\sum_{q,f,t} |\hat{Z}_q(f,t)|^2
}{\sum_{q,f,t}|X_q(f,t)|^2
},
\end{equation}
and averaged over time $T$, frequency $F$ and number of output channels $Q$.
The pattern is computed independently for each speech signal while keeping the input directivity pattern fixed.

Table \ref{tab:osn_results} provides results for the original FiLM-JNF architecture with 32~ms STFT windows \cite{Huang2025}, our adaptation to low-latency processing with 8~ms STFT windows, and the proposed \gls{FiLM-OSN}, evaluated on both loss functions.
    \begin{table}[t]
    \centering
      \caption{Comparison of low-latency implementations of FiLM-JNF and FiLM-OSN with the baseline regarding speech quality, intelligibility and distortions.}
    \resizebox{\linewidth}{!}{
    \begin{tabular}{ccccc}
         &Params& PESQ $\uparrow$ & ESTOI [\%] $\uparrow$ & SI-SDR [dB]$\uparrow$\\\hline
        Unprocessed & & $1.17\pm0.10$ &  $0.44\pm0.08$ & $-4.88\pm3.13$ \\\hline
        FiLM-JNF$_{\text{32ms}}$ &$950$~k& $2.10\pm0.46$& $0.74\pm0.08$ & $4.70\pm2.91$\\
    FiLM-JNF$_{\text{8ms}}$ & $950$~k& $1.72\pm0.36$  &  $0.66\pm0.10$ & $2.98\pm2.98$ \\
        FiLM-OSN$_{\text{8ms},\mathcal{L}_1}$ (ours) & $700$~k& $2.04\pm0.46$ & $0.77\pm0.08$  & $5.66\pm2.77$\\
        FiLM-OSN$_{\text{8ms},\mathcal{L}_{1+IPD}}$ (ours) &$700$~k& $2.06\pm0.46$& $0.77\pm0.08$ & $5.72\pm2.73$\\\hline
    \end{tabular}}
  
    \label{tab:osn_results}
\end{table}
For the FiLM-JNF$_{\text{8ms}}$, a performance drop is observed, which is attributable to the reduced sequence length of the wide-band \gls{LSTM}. Both \gls{FiLM-OSN} variants achieve superior performance relative to the low-latency baseline, indicating that the \gls{OSN} is more suitable for these conditions. Notably, training with the incorporated \gls{IPD} penalization results in marginal increase in \gls{PESQ} and \gls{SI-SDR} compared to the training on the $\mathcal{L}_1$. 

\begin{figure}[ht]
    \centering
    \includegraphics[width = 0.8\linewidth]{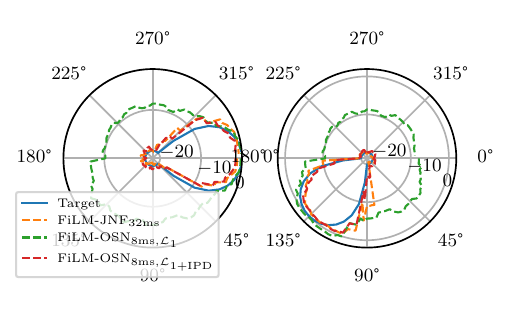}
    \caption{Estimated directivity patterns computed as the output-to-target power ratio averaged over 50 test speech signals spatialized using the KEMAR \gls{HARTF}.}
    
    \label{fig:beam_pattern_results}
\end{figure}

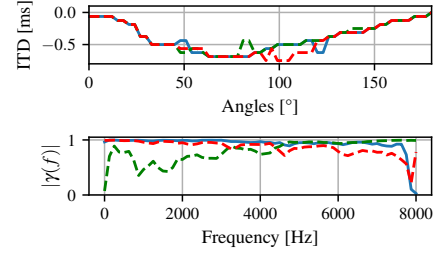
\begin{figure}[ht]
    \centering
    \includegraphics[width = 0.7\linewidth]{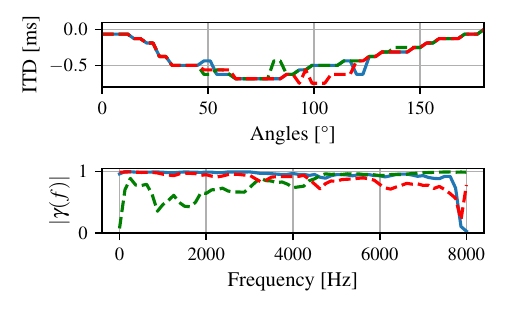}
    \caption{Comparison of ITD (top) and coherence (bottom) between FiLM-OSN$_{\text{8ms},\mathcal{L}_{1+IPD}}$ (\begin{tikzpicture}
        \fill[matplotlibred] (0,0) circle (0.3em);
            \end{tikzpicture}), FiLM-OSN$_{\text{8ms},\mathcal{L}_{1}}$ (\begin{tikzpicture} \fill[matplotlibgreen] (0,0) circle (0.3em);\end{tikzpicture}) and the target (\begin{tikzpicture}
                \fill[matplotlibblue] (0,0) circle (0.3em);
            \end{tikzpicture}).  Coherence is computed from a random test sample, and \gls{ITD} from a random spatialized speech sample.}
    \label{fig:cross_channel_coherence}
\end{figure}

Figure \ref{fig:beam_pattern_results} illustrates the estimated directivity patterns for the \gls{FiLM-JNF}$_\text{32ms}$ as well as the proposed FiLM-OSN variants, in comparison to the target directivity pattern. In particular, behavioral differences are observed between FiLM-OSN$_{\text{8ms},\mathcal{L}_1}$ and FiLM-OSN$_{\text{8ms},\mathcal{L}_{1+ \text{IPD}}}$.
While the FiLM-OSN$_{\text{8ms},\mathcal{L}_1}$ outperforms the baseline across all metrics, it entirely disregards the desired directivity pattern as depicted in Figure \ref{fig:beam_pattern_results}. 
A further analysis of the coherence, shown in Figure \ref{fig:cross_channel_coherence}, reveals distinctly different behavior of the FiLM-OSN$_{\text{8ms},\mathcal{L}_1}$ compared with the actual coherence. In contrast, FiLM-OSN$_{\text{8ms},\mathcal{L}_{1+ \text{IPD}}}$ exhibits a coherence behavior more closely aligned with the target, which suggests an improved preservation of cross-channel spectral coherence. Additionally, both approaches maintain the \gls{ITD} to a large extent, as evidenced by the \gls{ITD} comparison in Figure \ref{fig:cross_channel_coherence}.
These findings suggest that, due to the $\mathcal{L}_1$ time-domain loss, the \gls{FiLM-OSN} primarily learns to preserve the \gls{ITD}, while neglecting cross-channel coherence. Incorporating the \gls{IPD} enables the preservation of spectral relations, which leads to more accurate guidance towards the desired directivity pattern. 

\section{Conclusion}
In this article, we show that current neural directional filtering (NDF) approaches can be applied to hearing device scenarios. We propose a steerable low-latency NDF method for a behind-the-ear (BTE) hearing device setup, which allows users to adjust the focus in multi-talker scenarios. We also propose an additional phase difference deviation loss for our approach and show that this is crucial for achieving the desired directivity patterns. We suggest taking advantage of a state-space sequence-model architecture (Mamba) to achieve increased performance on the speech enhancement task under low-latency assumptions. The results on the diverse dataset show that our approach is also able to generalize to different microphone setups, since the training and testing set consists of various head diameters.

\bibliographystyle{IEEEbib.bst}
\bibliography{refs}

@STRING{TASLP={IEEE/ACM TASLP}}

@STRING{ICASSP={ICASSP}}

@STRING{IWAENC={IWAENC}}

@STRING{JASA={JASA}}

@STRING{Interspeech={Interspeech}}

@Article{Stone2003,
  author    = {Stone, Michael A. and Moore, Brian C. J.},
  journal   = {Ear and Hearing},
  title     = {Tolerable Hearing Aid Delays. III. Effects on Speech Production and Perception of Across-Frequency Variation in Delay},
  year      = {2003},
  issn      = {0196-0202},
  month     = apr,
  number    = {2},
  pages     = {175--183},
  volume    = {24},
  doi       = {10.1097/01.aud.0000058106.68049.9c},
  publisher = {Ovid Technologies (Wolters Kluwer Health)},
}

@article{Cherry1953,
    author = {Cherry, E. Colin},
    title = {Some Experiments on the Recognition of Speech, with One and with Two Ears},
    journal = JASA,
    volume = {25},
    number = {5},
    pages = {975-979},
    year = {1953},
    month = {09},
    issn = {0001-4966},
    doi = {10.1121/1.1907229},
    url = {https://doi.org/10.1121/1.1907229},
    eprint = {https://pubs.aip.org/asa/jasa/article-pdf/25/5/975/18731769/975_1_online.pdf},
}

@ARTICLE{QuanSpatial2024,

  author={Quan, Changsheng and Li, Xiaofei},

  journal=TASLP, 

  title={{SpatialNet}: Extensively Learning Spatial Information for Multichannel Joint Speech Separation, Denoising and Dereverberation}, 

  year={2024},

  volume={32},

  number={},

  pages={1310-1323},

  doi={10.1109/TASLP.2024.3357036}
        }

@ARTICLE{QuanOnSpatial2024,
author={Quan, Changsheng and Li, Xiaofei},
  journal={IEEE Signal Processing Letters}, 
  title={Multichannel Long-Term Streaming Neural Speech Enhancement for Static and Moving Speakers}, 
  year={2024},
  volume={31},
  number={},
  pages={2295-2299},
  doi={10.1109/LSP.2024.3418714}
}

@ARTICLE{Tesch2023,
  author={Tesch, Kristina and Gerkmann, Timo},
  journal=TASLP, 
  title={Insights Into Deep Non-Linear Filters for Improved Multi-Channel Speech Enhancement}, 
  year={2023},
  volume={31},
  number={},
  pages={563-575},
  doi={10.1109/TASLP.2022.3221046}}

@inproceedings{PandeyRNN2023,
  title     = {{A Simple RNN Model for Lightweight, Low-compute and Low-latency Multichannel Speech Enhancement in the Time Domain}},
  author    = {Ashutosh Pandey and Ke Tan and Buye Xu},
  year      = {2023},
  booktitle = Interspeech,
  pages     = {2478--2482},
  doi       = {10.21437/Interspeech.2023-2418},
  issn      = {2958-1796},
}

@article{Huang2025,
      title={Neural Directional Filtering with Configurable Directivity Pattern at Inference}, 
      author={Weilong Huang and Srikanth Raj Chetupalli and Emanuël A. P. Habets},
      year={2025},
      eprint={2510.20253},
      journal= {arXiv preprint arXiv:2510.20253},
      url={https://arxiv.org/abs/2510.20253}, 
}

@INPROCEEDINGS{Wechsler2024,
  author={Wechsler, Julian and Chetupalli, Srikanth Raj and Halimeh, Mhd Modar and Thiergart, Oliver and Habets, Emanuël A. P.},
  booktitle=IWAENC, 
  title={Neural Directional Filtering: Far-Field Directivity Control with a Small Microphone Array}, 
  year={2024},
  volume={},
  number={},
  pages={459-463},
  doi={10.1109/IWAENC61483.2024.10693965}}

@ARTICLE{Gu2024,
  author={Gu, Rongzhi and Luo, Yi},
  journal=TASLP, 
  title={{ReZero}: Region-Customizable Sound Extraction}, 
  year={2024},
  volume={32},
  number={},
  pages={2576-2589},
  doi={10.1109/TASLP.2024.3393713}}

@ARTICLE{Wang2023,
  author={Wang, Zhong-Qiu and Wichern, Gordon and Watanabe, Shinji and Le Roux, Jonathan},
  journal=TASLP, 
  title={{STFT}-Domain Neural Speech Enhancement With Very Low Algorithmic Latency}, 
  year={2023},
  volume={31},
  number={},
  pages={397-410},
  doi={10.1109/TASLP.2022.3224285}}

@ARTICLE{Mesgarani2019,
  author={Luo, Yi and Mesgarani, Nima},
  journal=TASLP, 
  title={{Conv-TasNet}: Surpassing Ideal Time–Frequency Magnitude Masking for Speech Separation}, 
  year={2019},
  volume={27},
  number={8},
  pages={1256-1266},
  doi={10.1109/TASLP.2019.2915167}}

@article{Denk2018,
    author = {Florian Denk and Stephan M. A. Ernst and Stephan D. Ewert and Birger Kollmeier},
    title ={Adapting Hearing Devices to the Individual Ear Acoustics: Database and Target Response Correction Functions for Various Device Styles},
    journal = {Trends in Hearing},
    volume = {22},
    number = {},
    year = {2018},
    doi = {10.1177/2331216518779313},
}

@Misc{Pausch2022,
  author    = {Pausch, Florian and Doma, Shaimaa Ahmed Bassim Ali and Fels, Janina},
  title     = {{IHTA-indHARTF} - Database of individual behind-the-ear hearing-aid-related transfer functions with high spatial resolution},
  year      = {2022},
  doi       = {10.18154/RWTH-2022-04267},
  journal   = {doi:10.18154/RWTH-2022-04267},
  language  = {en},
  publisher = {RWTH Aachen University},
}

@Article{Perez2018,
  author    = {Perez, Ethan and Strub, Florian and De Vries, Harm and Dumoulin, Vincent and Courville, Aaron},
  journal   = {Proceedings of the AAAI Conference on Artificial Intelligence},
  title     = {FiLM: Visual Reasoning with a General Conditioning Layer},
  year      = {2018},
  issn      = {2159-5399},
  month     = apr,
  number    = {1},
  volume    = {32},
  doi       = {10.1609/aaai.v32i1.11671},
  publisher = {Association for the Advancement of Artificial Intelligence (AAAI)},
}

@article{welker2025,
      title={Real-Time Streamable Generative Speech Restoration with Flow Matching}, 
      author={Simon Welker and Bunlong Lay and Maris Hillemann and Tal Peer and Timo Gerkmann},
      year={2025},
      journal = {arXiv preprint arXiv:2512.19442},
      url={https://arxiv.org/abs/2512.19442}, 
}

@INPROCEEDINGS{Wu2025,
  author={Wu, Haibin and Braun, Sebastian},
  booktitle=ICASSP, 
  title={Ultra-Low Latency Speech Enhancement -A Comprehensive Study}, 
  year={2025},
  doi={10.1109/ICASSP49660.2025.10889823}}

@ARTICLE{Tesch2024,
  author={Tesch, Kristina and Gerkmann, Timo},
  journal=TASLP, 
  title={Multi-Channel Speech Separation Using Spatially Selective Deep Non-Linear Filters}, 
  year={2024},
  volume={32},
  number={},
  pages={542-553},
  doi={10.1109/TASLP.2023.3334101}}

@ARTICLE{Westhausen2024,
  author={Westhausen, Nils L. and Kayser, Hendrik and Jansen, Theresa and Meyer, Bernd T.},
  journal=TASLP, 
  title={Real-Time Multichannel Deep Speech Enhancement in Hearing Aids: Comparing Monaural and Binaural Processing in Complex Acoustic Scenarios}, 
  year={2024},
  volume={32},
  number={},
  pages={4596-4606},
  doi={10.1109/TASLP.2024.3473315}}

@incollection{Habets2010,
author="Habets, Emanu{\"e}l A. P.
and Benesty, Jacob
and Gannot, Sharon
and Cohen, Israel",
title="The {MVDR} Beamformer for Speech Enhancement",
bookTitle="Speech Processing in Modern Communication: Challenges and Perspectives",
year="2010",
publisher="Springer",
address="Berlin, Heidelberg",
pages="225--254",
isbn="978-3-642-11130-3",
doi="10.1007/978-3-642-11130-3_9",
url="https://doi.org/10.1007/978-3-642-11130-3_9"
}

@incollection{Wiley1999,
author = "Kanti V. Mardia and Peter E. Jupp",
publisher = {John Wiley \& Sons, Ltd},
isbn = {9780470316979},
title = {Summary Statistics},
booktitle = {Directional Statistics},
chapter = {2},
pages = {13-24},
doi ={https://doi.org/10.1002/9780470316979.ch2},
year = {1999}
}

@INPROCEEDINGS{Pandey2022,
  author={Pandey, Ashutosh and Xu, Buye and Kumar, Anurag and Donley, Jacob and Calamia, Paul and Wang, DeLiang},
  booktitle=ICASSP, 
  title={Multichannel Speech Enhancement Without Beamforming}, 
  year={2022},
  volume={},
  number={},
  pages={6502-6506},
  doi={10.1109/ICASSP43922.2022.9746704}}

@inproceedings{Gu2023,
  author    = {Gu, Albert and Dao, Tri},
  title     = {Mamba: Linear-Time Sequence Modeling with Selective State Spaces},
  year      = {2023},
  booktitle = {arXiv preprint arXiv:2312.00752},
  copyright = {Creative Commons Attribution 4.0 International},
  doi       = {10.48550/ARXIV.2312.00752}
}

@INPROCEEDINGS{RixPESQ2001,
  author={Rix, A.W. and Beerends, J.G. and Hollier, M.P. and Hekstra, A.P.},
  booktitle=ICASSP, 
  title={Perceptual evaluation of speech quality {(PESQ)} -- a new method for speech quality assessment of telephone networks and codecs}, 
  year={2001},
  volume={2},
  number={},
  pages={749-752 vol.2},
  doi={10.1109/ICASSP.2001.941023}}

@INPROCEEDINGS{Thiergart2012,
  author={Thiergart, Oliver and Habets, Emanuel A. P.},
  booktitle=IWAENC, 
  title={Sound Field Model Violations in Parametric Spatial Sound Processing}, 
  year={2012}
}

\end{document}